# DEEP CNN FRAMEWORKS COMPARISON FOR MALARIA DIAGNOSIS

Priyadarshini Adyasha Pattanaik, Zelong Wang and Patrick Horain

*Télécom SudParis, Institut Polytechnique de Paris*
*9 rue Charles Fourier, 91011 ÉvryCedex, France*
ipriyadarshinipattanaik@gmail.com, zelong.wang@ telecom-sudparis.eu, Patrick.Horain@telecom-sudparis.eu

**Abstract**

We compare Deep Convolutional Neural Networks (DCNN) frameworks, namely AlexNet and VGGNet, for the classification of healthy and malaria-infected cells in large, grayscale, low quality and low resolution microscopic images, in the case only a small training set is available. Experimental results deliver promising results on the path to quick, automatic and precise classification in unstrained images.

## 1 Introduction

Malaria is a key cause of infectious death worldwide causing 219 million infections and 435,000 deaths in 2017 according to the World Health Organization [1]. In most instances, malaria can only be diagnosed by manually checking microscopic slides [2]. Microscopic blood smear images imaging is essential to cellular analysis as the morphology and texture characteristics of erythrocyte cells changes during malaria infection and progression. Without staining, however, red blood cells and parasites have close transparency, resulting in low contrast. Deep learning helps to enhance the low-resolution image and restore high-resolution image correctly, by inferring high-frequency components of a low-resolution image. [6].

In this paper, we examine the abilities of some classical deep convolutional neural networks, including AlexNet and VGGNet [7], to improve the classification accuracy when trained on a small and complex unstained image dataset.

The rest of the manuscript is structured as follows. In Section II we present the DCNN framework, while in Section III we describe the various methods and their results involved. Section IV states our experimental results with discussion and Section V carries the conclusion.

## 2 DCNN Framework

A convolutional neural network (CNN) [6][7] is a feed-forward process for image recognition. A typical CNN [4] is characterized by four layers i.e. local connections, activation, pooling layers, and dense layers. Deep CNNs of 5 convolutional layers with different size filters, i.e. AlexNet successfully delivered multi-categorical classification and have created a breakthrough boost in the application of CNNs. Following this initial success, VGGNet consists of 13 convolutional layers and a unique size filter of $3 \times 3$, won the localization task in ILSVRC 2014. Both the deep learning CNNs use dropout technique to prevent overfitting.

## 3 Methods

Our dataset consists of four unstained images of 2456×2054 pixels in 16-bit grayscale, captured from axial LED illumination. The datasets are not completely labelled missing few annotations and include overlapping cells.

One of the input images is shown in Figure 1. Each input image contains hundreds of possibly infected cells along with artifact and platelets.

### 3.1 Pre-Processing

Before feeding the CNN, the original large images are split into 71 × 71 pixels. A Gaussian blur and circular Hough transform [8] allows selecting those tiles that hold complete blood cells. In our experiments, we achieved a cell detection accuracy (FP+FN / TP+FP) in the range of 95.5 % to 98.2 %.

### 3.2 Segmentation

Next, tiles are annotated by hand as either Infected or Healthy as we have small training set, especially for infected cells we augment the tile dataset by rotation and flip. At last, images are normalized by subtracting the mean of the entire dataset.

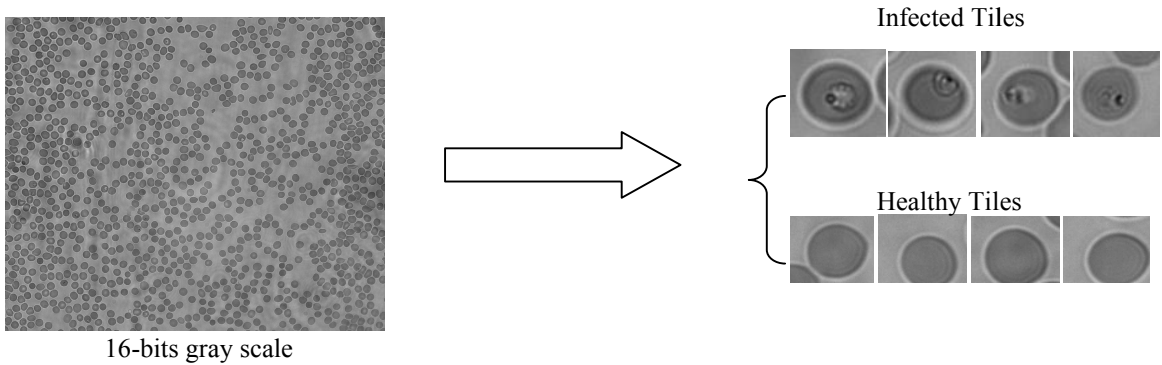

16-bits gray scale

**Figure. 1 Slicing original images into tiles of equal size containing the whole cell**

### 3.3 DCNN Framework Training

In order to train and evaluate our proposed framework, we implement five-fold cross-validation on the whole 14,740 tiles of each 71×71 pixels, where 80% of the tiles are used for training and rest 20% for testing. We used transfer learning on two CNN models, AlexNet and VGGNet respectively, with a powerful GPU of GeForce RTX 2080 Ti to accelerate the training. The CNN models were trained to 100 epochs with 0.0001 learning rate, dropout rate 0.5 and batch size of 128.

## 4 Experimental Results and Discussion

This section describes the different aspects of the effectiveness of our proposed DCNN framework and evaluates the procedure using various evaluation metrics. Effectiveness of the proposed deep learning framework is obtained in terms of image retrieval feature extractor representations. Figure 2 represents the feature maps obtained from the convolutional layers and filters of AlexNet (Figure 2 (a), (b))and VGGNet (Figure 2 (c), (d)) respectively. Quality qualification performed by the proposed framework is measured in terms of sensitivity, specificity, and accuracy. We present a conceptual baseline comparison of DCNN framework with different existing deep learning techniques.

We determined the epoch of trained model according to the training curve in Figure 3, represents the loss and accuracy balance of training dataset and validation dataset in order to prevent overfitting, selecting different epoch for different training. In Figure 3, each case contains two curves of loss and accuracy representing the blue line as train dataset and red line as validation dataset. We can see that losses decrease and accuracies increase as per the epoch count, however after ~10 epochs, the validation accuracy reaches to constant. According to the epoch selected, VGGNet uses fewer epochs than AlexNet to reach a balance that represents training dataset faster. However, the training time of VGGNet is a bit higher as compared to AlexNet due to the number of layers and filters involved for training the large quantity of the dataset.

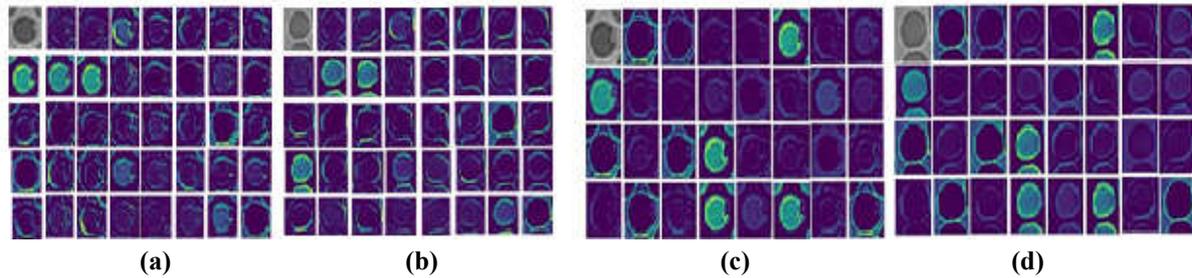

(a)　　　　　　　　　(b)　　　　　　　　　(c)　　　　　　　　　(d)

**Figure 2. Feature map outputs of the convolutional layers of AlexNet for (a) malaria-infected and (b) healthy cell visualization. Feature map outputs of the convolutional layers of VGGNet for (c) malaria-infected and (d) healthy cell visualization.**

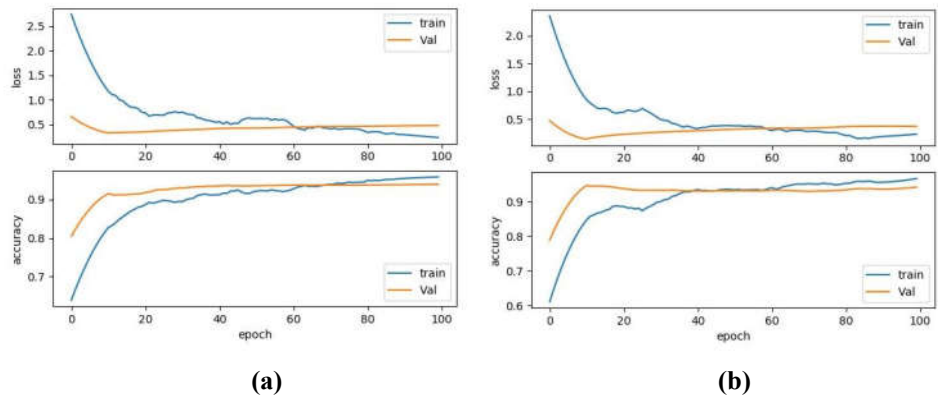

(a)　　　　　　　　　(b)

**Figure 3. Comparison of the convergence speed of validation and training dataset of (a) AlexNet and (b) VGGNet w.r.t loss and accuracy.**

The most remarkable success of this work is that the best DCNN framework proposed with pre-processing in this research for malaria parasite detection has a low false-negative rate of 4% in AlexNet and 3% in VGGNet. The main objective of the work is to reduce the false-negative without missing any parasite infected cell. This result is significantly better than the rest of the existing deep learning networks in the area of malaria detection. The proposed framework includes the AlexNet and achieves an F-score of 95.8%, specificity of 95.6% and sensitivity of 96.5. The VGGNet delivers an F-score of 96.5%, specificity of 96.5% and sensitivity of 96.8%.

**Table 1. Comparison between the proposed deep neural framework with other deep learning techniques**

| Models | Specificity(TNR %) | Sensitivity(TPR %) | F- score (%) |
|---|---|---|---|
| Deep Belief Networks [5] | 95.9 | 97.6 | 89.6 |
| Stacked Autoencoder [7] | 97.4 | 72.3 | 78.4 |
| LeNet [7] | 95.2 | 89.2 | 82.1 |
| **AlexNet [ Proposed DCNN]** | **95.6** | **96.5** | **95.8** |
| **VGGNet[ Proposed DCNN]** | **96.5** | **96.8** | **96.5** |

**Table 2. Compared performance (AlexNet and VGGNet) with different training options**

| Parameters (Performance Measures) | MODELS | |
| | AlexNet | VGGNet |
|---|---|---|
| Training Time/40 epoch | 21 minutes 40 seconds | 1 hour 5 minutes 30 seconds |

| Test Time | 3 seconds | 9 seconds |
|---|---|---|
| Accuracy | 96 | 96.7 |
| Sensitivity(TPR %) | 96.5 | 96.8 |
| Specificity(TNR %) | 95.6 | 96.5 |
| F- score (%) | 95.8 | 96.5 |
| Tile Images carrying complete cells (71 × 71 Pixels)<br>Healthy　　　　　Infected<br>Train: 632　　　　Train: 568<br> Val: 104　　　　　Val: 96<br>Test: 320　　　　Test: 288 | | |

## 5  Conclusions

In this work, we have compared AlexNet and VGGNet frameworks to classify blood smear image tiles into 2 classes, including 3 different blood cells. Through comparison of CNNs with different internal depths, it shows that VGGNet is better at extracting, which improves 3% false-negative prediction rate than AlexNet. We plan to expand the dataset by including other imaging modalities on the classification process.

## Acknowledgments

This work was jointly financed by the Carnot Institutes M.I.N.E.S and TSN, France. Special thanks go to Pitié-Salpêtrière Hospital, Medical Biology, and Pathologies team, for preparing and digitizing the slides and to TRIBVN HealthCare for sharing images.